\documentclass[prl,showpacs,twocolumn]{revtex4}%
\usepackage{epsf}
\usepackage{amsmath}
\usepackage{graphicx}
\usepackage{amsfonts}
\usepackage{amssymb}%

\begin{document}
\title{Bell inequality violation with two remote atomic qubits}
\date{\today } 
\author{D. N. Matsukevich}
\altaffiliation[Electronic address: ]{dmats@umd.edu}
\author{P. Maunz}
\author{D. L. Moehring}
\altaffiliation[Present address: ]
{Max-Planck-Institut f\"{u}r Quantenoptik, Hans-Kopfermann-Str. 1, 
D-85748 Garching, Germany} 
\author{S. Olmschenk}
\author{C. Monroe} 
\affiliation{Department of Physics and Joint Quantum Institute,
University of Maryland, College Park, Maryland, 20742} 
\begin{abstract}
We observe violation of a Bell inequality between the quantum states
of two remote Yb$^+$ ions separated by a distance of about one meter
with the detection loophole closed. The heralded entanglement of two
ions is established via interference and joint detection of two
emitted photons, whose polarization is entangled with each ion. The
entanglement of remote qubits is also characterized by full quantum
state tomography.
\end{abstract}
\maketitle

In 1964 Bell showed that in all local realistic theories, correlations
between the outcomes of measurements in different parts of a physical
system satisfy a certain class of inequalities
\cite{bell}. Furthermore, he found that some predictions of quantum
mechanics violate these inequalities. Starting with the first
experimental tests of Bell inequalities with photons
\cite{freedman,aspect,weihs}, violation of a Bell inequality has
been observed in a wide range of systems including protons
\cite{bellproton}, K-mesons \cite{bellkaon}, ions \cite{wineland},
neutrons \cite{bellneutron}, B-mesons \cite{bellmeson}, heterogeneous
atom-photon systems \cite{david2,mats2} and atomic ensembles
\cite{mats,chou}. Demonstration of the violation of a Bell inequality
has also become a routine technique to verify the presence of
entanglement and check the security of a quantum communication link
\cite{gisin}.

In order to exclude all local realistic theories, a rigorous
experimental test of a Bell inequality must satisfy two
conditions. First, the measurement time has to be sufficiently short
such that no information traveling at the speed of light can propagate
from one qubit to another during the measurement (locality loophole).
Second, the efficiency of the quantum state detection has to be high
enough such that it is impossible to mimic a Bell inequality violation
by selective choice of the successful measurement events (detection
loophole). Since photons can propagate over a long distance and be
detected fast, the locality loophole was first closed in a photon
system \cite{aspect,weihs}.  On the other hand, high detection
efficiency and deterministic preparation of an entangled state of
trapped ions has closed the detection loophole in a system of two
trapped ions separated by $\sim 3\: \mu$m \cite{wineland}. Although
several experimental schemes for a loophole-free Bell inequality test
have been proposed \cite{kwiat94,huelga,fry,garcia,simon}, no
experiment to date has closed both loopholes simultaneously.

One of these proposals (by Simon and Irvine \cite{simon}) combines the
advantages of photons and trapped ions. This protocol starts by
preparing two spatially separated ions, each entangled with its
emitted photon.  These photons are then sent to an intermediate
location where a partial Bell state analysis is performed. Successful
detection of an entangled state of two photons unambiguously heralds
the preparation of an entangled state of the two ions. The Bell
inequality violation is then verified by local rotation and detection
of the ion qubits \cite{wineland}.

In this Letter, we report an important step towards implementation of
this protocol with the observation of a Bell inequality violation
using two $^{171}\rm{Yb}^+$ ions separated by about $1$ meter.  In
contrast to our previous work where the photonic qubit was encoded in
the frequencies of a photon \cite{david1,davidjosa}, here we use the
polarization degree of freedom for the photonic qubit and two nearly
degenerate states of the atom to encode the atomic qubit. This allows
for measurement of both atomic and photonic qubits in arbitrary bases
and for characterization of the generated ion-photon and ion-ion
entangled states by quantum state tomography, resulting in an ion-ion
entanglement fidelity of 81\%.  Together with the high efficiency of
detecting the quantum state of an ion, such a fidelity makes it
possible to observe a Bell inequality violation between two distant
particles with the detection loophole closed. With an even larger
separation between the ions or faster detection of the atomic qubit,
the technique demonstrated here may ultimately allow for a
loophole-free Bell inequality test \cite{simon,weinfurter}.

\begin{figure}
\centerline{
\epsfxsize=8cm
\epsfbox{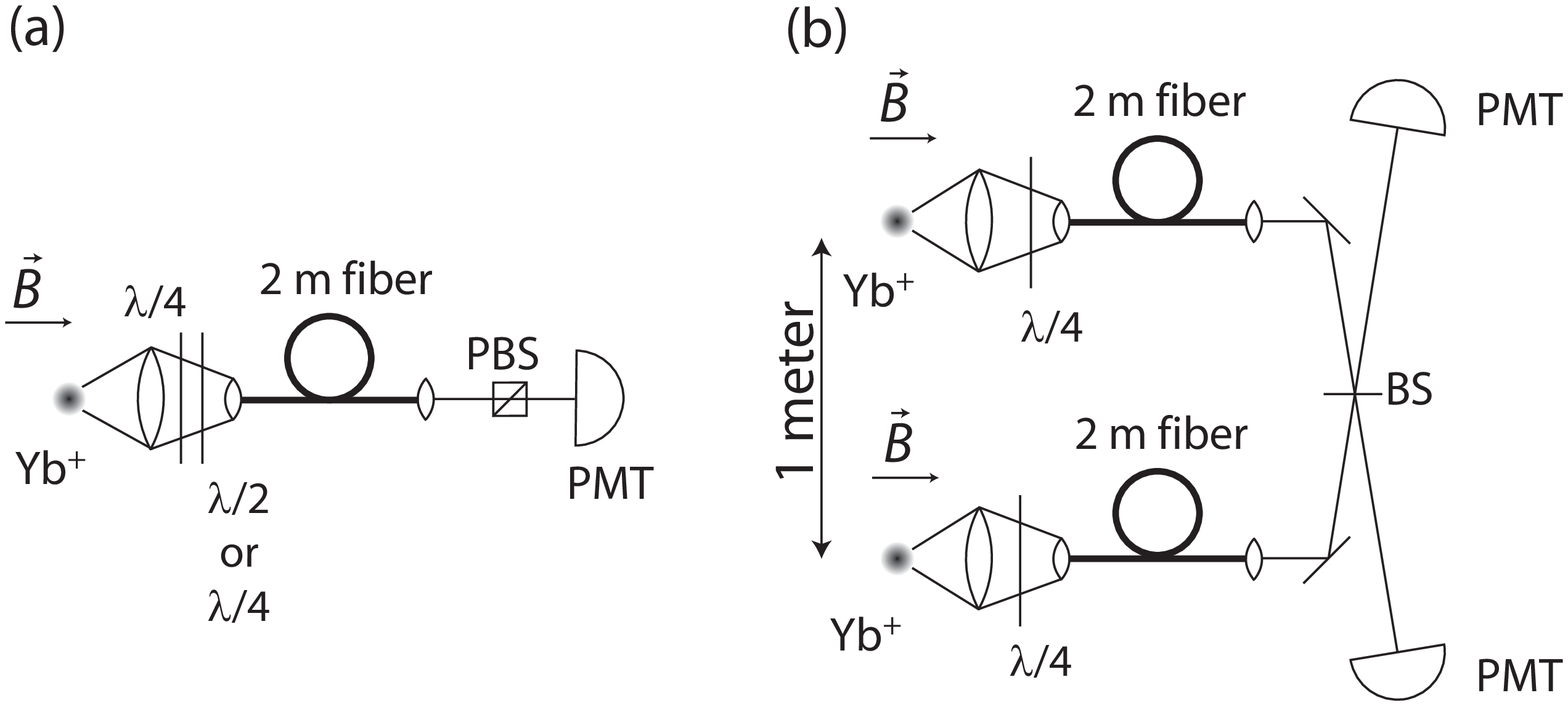}}
\caption{\label{fig:setup}Experimental setup for (a) the ion-photon
and (b) ion-ion experiments. $\vec B$ shows the direction of
applied magnetic field. PBS: polarizing beamsplitter; PMT:
photomultiplier tube; BS: 50/50 nonpolarizing beamsplitter;
$\lambda/2$: half-wave plate; $\lambda/4$:\ quarter-wave plate.}
\end{figure}

The experimental setup is shown in Fig.~\ref{fig:setup}. A single
$^{171}$Yb$^{+}$ ion is stored in each of two rf-Paul traps located in
independent vacuum chambers. The ions are placed in a magnetic field
of 4.6 G parallel to the direction of the quantization axis. A 700 ns
light pulse at 369.5 nm resonant with the ${^{2}S_{1/2},F = 1}
\rightarrow {^{2}P_{1/2},F = 1}$ transition optically pumps the ions
in both traps to the $^{2}S_{1/2}, F = 0, m_F=0$ $(^{2}S_{1/2}|0,
0\rangle)$ state.  The $^{2} P_{1/2}$ state has a probability of
$\simeq 0.005$ to decay to the metastable $^{2} D_{3/2}$ state. To
prevent population trapping in this state, the ion is illuminated with
935.2 nm light resonant with the $^{2}D_{3/2} \rightarrow
{^3D[3/2]_{1/2}}$ transition \cite{steve}.  After optical pumping, a
$\simeq 2$~ps light pulse from a frequency doubled, mode-locked,
Ti-sapphire laser, polarized linearly along the direction of the
magnetic field, transfers the population to the excited
$^{2}P_{1/2}|1,0\rangle$ state with near unit efficiency. Since the
duration of the excitation pulse is much shorter than the excited
state lifetime ($\simeq 8$ ns), at most one 369.5 nm photon can be
emitted by the ion \cite{peter}. The duration of each optical pumping
and excitation cycle is 1.4~$\mu s$. After 107 cycles, the ions are
Doppler cooled for 40~$\mu s$. The average overall excitation rate is
0.52~MHz.

When viewed along the quantization axis, the $^{2}P_{1/2}|1,0\rangle$
state either decays to the $^2S_{1/2}| 1,1 \rangle$ state emitting
a left circular ($\sigma^{-}$) polarized photon or to the $^2S_{1/2}|
1, -1 \rangle$ state emitting a right circular ($\sigma^{+}$)
polarized photon. According to the dipole radiation pattern, $\pi$
polarized photons emitted due to decay to the $^2S_{1/2}| 0,0 \rangle$
state cannot propagate along the quantization axis
direction. Therefore, when the photon is emitted along the
quantization axis, the state of each ion is entangled with the
polarization state of its emitted photon:
\begin{equation}
\Psi = \frac{1}{\sqrt 2} 
 ( | 1, 1 \rangle | \sigma^{-} \rangle 
 - | 1, -1 \rangle | \sigma^{+} \rangle).
\end{equation}
For each ion, the emitted photons are collected by an imaging lens
(numerical aperture 0.23) and sent through a $\lambda / 4$ wave plate to
convert $\sigma^{+}$ or $\sigma^{-}$ circular polarization to linear
horizontal ($H$) or vertical ($V$) polarization, respectively. The
state of each ion-photon system given the photon passes the
quarter-wave plate can be written as
\begin{equation}
 \Psi = \frac{1}{\sqrt 2} 
  (     | 1,  1 \rangle | V \rangle  - i | 1, -1 \rangle | H \rangle).
\end{equation} 
The photons from each ion are coupled to a single mode fiber to
facilitate mode-matching on a nonpolarizing 50/50
beamsplitter. Photons at the output ports of the beamsplitter are
detected with photomultiplier tubes (PMTs) (see
Fig.~\ref{fig:setup}(b)). The contrast of interference between two
modes is $97 \% $, the quantum efficiency of each PMT is about 15\%,
and the count rate due to the dark counts and background light leakage
is about 3 Hz. The arrival times of the photo-electric pulses from the
PMTs are recorded by a time to digital converter.  Coincidence
detection of two photons interrupts an experimental time sequence and
triggers a sequence of microwave pulses to perform rotation of the
ion qubits, followed by state detection of the ions using standard
fluorescence techniques \cite{steve}.

Given perfect mode-matching of the input single photon wavepackets on
the beamsplitter, detection of a photon at each output port of the
beamsplitter corresponds to a successful measurement of photons in the
state \cite{mandel}
\begin{equation}
\Psi_{ph} = \frac{1}{\sqrt{2}}(| H \rangle | V \rangle - | V \rangle
| H \rangle).
\end{equation}
This projects ions ($a$) and ($b$) onto the entangled state
\cite{simon,davidjosa}
\begin{equation}
\Psi_{ion} = \frac{1}{\sqrt{2}}(| 1,1  \rangle_a | 1, -1 \rangle_b 
- |1, -1 \rangle_a | 1, 1  \rangle_b ).
\label{eq:ion_entangle}
\end{equation}

Due to the Zeeman splitting of the ground states of the ion, the
emitted $\sigma^{+}$ and $\sigma^{-}$ polarized photons have a
frequency difference of about 13 MHz. Nevertheless, it is still
possible to get the entangled state of Eq.(\ref{eq:ion_entangle}) if
the photons with the same polarization also have the same
frequency. 

Quantum state tomography of the atomic qubits requires the ability to
detect the state of each ion in an arbitrary basis. For this we first
apply a resonant microwave $\pi$ pulse to transfer the population from
$^{2}S_{1/2}| 1, -1 \rangle$ to the $^{2}S_{1/2}| 0, 0 \rangle$
state. Next, a second microwave pulse resonant with the
$^{2}S_{1/2}|0, 0\rangle \leftrightarrow {^{2}S_{1/2}}| 1, 1 \rangle$
transition, with a controlled duration and phase relative to the first
pulse, is applied to perform a qubit rotation (see Fig.~2).  As a
result, the state of the ion is transformed as follows:
\begin{eqnarray}
 \cos (\theta_i / 2) | 1, 1 \rangle + \sin (\theta_i / 2) e^{i \phi} 
|1, -1 \rangle & \rightarrow & |1, 1\rangle \nonumber \\ 
- \sin (\theta_i / 2) e^{- i \phi} | 1, 1 \rangle + \cos (\theta_i / 2) 
|1, -1 \rangle & \rightarrow & |0, 0\rangle.
\end{eqnarray}
Here $\theta_i$ is proportional to the duration of the second microwave
pulse, and $\phi$ is the relative phase between the first and second
pulses \cite{david2,david1,blinov}.
\begin{figure}
\centerline{
\epsfxsize=7cm
\epsfbox{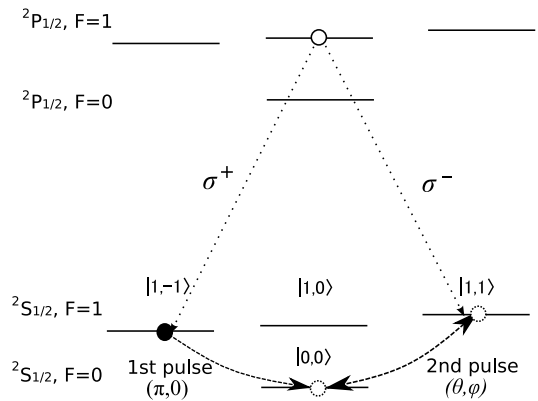}}
\caption{The ion-photon entanglement scheme and sequence of microwave
pulses for ion qubit manipulations. $\sigma^{+}$ and $\sigma^{-}$ are
right and left circular polarization of a photon, $\theta$ is
proportional to the duration of the second microwave pulse and $\phi$
is a relative phase between first and second microwave pulses.}
\end{figure}

The fluctuations of the magnetic field at the position of the ion can
change the phase that each ion acquires before state detection.
To keep the magnitude of the magnetic field constant, the
experimental sequence is interrupted about 20 times a second to
perform a Ramsey experiment. The ions in each trap are first optically
pumped to the $| 0, 0 \rangle$ ground state. Two microwave $\pi/2$
pulses resonant with the $|0, 0 \rangle \leftrightarrow | 1, -1
\rangle$ transition separated by 200 $\mu s$ are applied and then the
state of each ion is detected. The probability to find the ion in
the $|1,-1\rangle$ state is continuously monitored and the current in
the bias coil is adjusted to keep the magnetic field magnitude
constant. We estimate that fluctuations of the magnetic field do not
exceed 1 mG over the several days of experiment.

Following microwave rotations, the state of the Yb$^{+}$ ion is
detected. The 369.5~nm light resonant with the ${^2}S_{1/2}, F = 1
\rightarrow {^2}P_{1/2}, F = 0$ transition impinges on the ion and the
fluorescence is detected with a PMT. Ideally, if an ion is in the $F =
1$ state, it scatters this light. On the other hand, if the ion is in
the $F = 0$ state, it remains dark, allowing the quantum state of the
atomic qubit to be distinguished with an efficiency of about 98\%
\cite{steve}. It is important to note that unlike single photon
detection, every attempt to detect the state of an ion gives a
result. The efficiency quoted here is the probability that this result
is correct.

To verify that the emitted photon is indeed entangled with the ion, we
temporarily add an additional half-wave or quarter-wave plate and a
polarizer (see Fig.~\ref{fig:setup}a). In this case the ion
manipulation and detection sequence is triggered on the detection of a
single photon.
\begin{figure}
\centerline{
\epsfxsize=9.5cm
\epsfbox{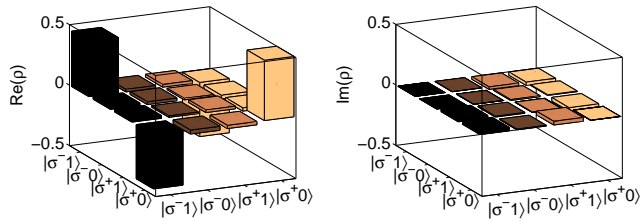}
}
\caption{\label{fig:ion_photon_state}Real (left) and imaginary (right)
parts of reconstructed density matrix for the system of a single
photon and an ion. Results are based on 42495 events. }
\end{figure}

Quantum state tomography is performed for full characterization of the
the state of the ion-photon system \cite{weinfurter,wilk}.  We chose
to measure both the ion and the photon states in the $\{\sigma_j \
\sigma_k; j,k = x,y,z \}$ bases.  Each measurement is integrated for
100 seconds with an average rate of about 25 ion-photon entanglement
events per second. The state tomography algorithm follows the maximum
likehood estimation technique described in \cite{kwiat}, with the
result shown in Fig.~\ref{fig:ion_photon_state}. From the
reconstructed density matrix, we calculate the entanglement fidelity
$F_{ip} = 0.925 \pm 0.003$, concurrence $C_{ip} = 0.861 \pm 0.006$ and
entanglement of formation $E_{Fip} = 0.805 \pm 0.008$. We also have
measured a Bell inequality parameter $S$ for our ion-photon system
\cite{david2}. The result of the measurement ($S = 2.54 \pm 0.02 > 2$)
clearly violates the Clauser-Horne-Simony-Holt version of the Bell
inequality \cite{chsh}, described below.

This high measured entanglement fidelity between a single ion and a
single photon allows us to establish entanglement between two remote
ions in violation of a Bell inequality. With two ions simultaneously
excited, the photoelectric pulses from the PMTs on both output ports of
a beamsplitter arriving within a $\pm 25$ ns coincidence window
indicate a successful entanglement event. Following the second
photoelectric pulse from the PMTs, the states of both ions are rotated
and detected as described above.

To verify the Bell inequality violation, we keep the phase $\phi$ for
both ions at $0$ and vary $\theta$.  Following
Clauser-Horne-Simony-Holt (CHSH) \cite{chsh}, we calculate the
correlation function $E(\theta_a, \theta_b)$ given by
\begin{equation}
E(\theta_a, \theta_b) = 
  p(\theta_a, \theta_b) + p(\theta_a^{\bot}, \theta_b^{\bot}) 
- p(\theta_a^{\bot}, \theta_b) - p(\theta_a, \theta_b^{\bot}),
\end{equation} 
where $p(\theta_a, \theta_b)$ is the probability to find ion ($a$) in
the state $ \cos (\theta_a / 2) |1,1\rangle + \sin (\theta_a / 2)
|1,-1\rangle $ and ion ($b$) in the state $ \cos (\theta_b / 2)
|1,1\rangle + \sin (\theta_b / 2) |1,-1\rangle $, $\theta_{a,b}^{\bot}
= \theta_{a,b} + \pi$.

The CHSH version of a Bell inequality states that for all local
realistic theories
\begin{equation}
S =| E(\theta_a, \theta_b) + E(\theta'_a, \theta_b) |
  + | E(\theta_a, \theta'_b) - E(\theta'_a, \theta'_b) | \leq 2.
\end{equation}
The result of the Bell inequality measurement is given in Table
\ref{tab:ion_ion}, based on 2276 coincidence events.  On average, we
observed 1 entanglement event per 39 sec.  The result $S = 2.22 \pm
0.07$ represents a Bell inequality violation by more than three
standard deviations. Since every heralded entanglement event is
followed by the measurement of the qubit states, the Bell inequality
violation is observed with the detection loophole closed.
\begin{figure}
\centerline{
\epsfxsize=9.5cm
\epsfbox{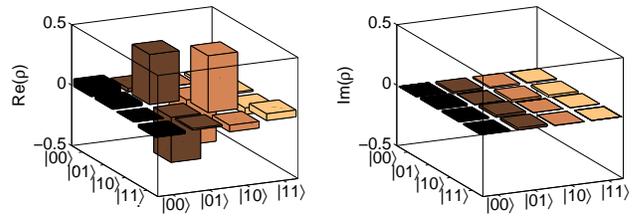}
}
\caption{\label{fig:ion_ion} Real (left) and imaginary (right) parts
of reconstructed density matrix for the system of two entangled
ions. Results are based on 2121 coincidence events.}
\end{figure}
\begin{table}
\caption{\label{tab:ion_ion} Measured correlation function $E(\theta
_a, \theta _b)$ and CHSH parameter $S$ for the ion-ion state. Errors
are based on the statistics of the photon counting events.}
\begin{ruledtabular}
\begin{tabular}{ccccc}
$\theta _a$ & $\theta _b $& $E(\theta_a, \theta _b)$  \\
\hline
$\pi/2$  & $\pi/4$      &  $-0.518  \pm 0.036$   \\
$\pi/2$  & $3 \pi /4$   &  $-0.546  \pm 0.034$   \\
0        & $\pi/4$      &  $-0.581  \pm 0.034$   \\
0        & $3 \pi /4$   &  $0.573 \pm 0.035$   \\
         &              &  $S=2.22 \pm 0.07$    \\
\end{tabular}
\end{ruledtabular}
\end{table}

We also performed state tomography for the entangled state of the two
ions. As in the ion-photon case, ion measurements were performed in
the $\{\sigma_j \ \sigma_k; j,k = x,y,z \}$ bases, 
with the result shown in Fig.~\ref{fig:ion_ion}.
From this density matrix we estimate the entanglement fidelity $F_{ii}
= 0.813 \pm 0.015$, concurrence $C_{ii} = 0.64 \pm 0.03$, and
entanglement of formation $E_{Fii} = 0.52 \pm 0.04$. 

The ion-photon entanglement fidelity is mainly limited by the
detection efficiency of the ion state (3\% decrease of fidelity), a
fluctuating ambient magnetic field that causes ion dephasing (2\%),
imperfect compensation of the stress-induced birefringence in the
viewports of our vacuum chambers and imperfections in polarization
control for light propagating through the fibers (1\%), decay of the
ion to the $m_F = 0$ state due to nonzero solid angle of the photon
collection (1\%), and PMTs dark counts ($ < 0.5 \%$).  In addition,
interference contrast of the interferometer contributes to the reduced
entanglement fidelity of two ions (9 $\pm$ 3)\% compared to the ideal
interference between the photons $F_{ii}^{ideal} = 86\%$, calculated
from the reconstructed ion-photon state. The 13 fold improvement in
the entanglement generation rate compared to our previous experiment
\cite{david1} is due to a different excitation scheme that allows the
transfer of all the population to the excited state and a different
direction for photon collection with respect to the applied magnetic
field that does not require polarization filtering of the collected
photon.

Here we have successfully extended the separation between entangled
particles by more than 5 orders of magnitude, as compared to the
previous Bell inequality test performed with the detection loophole
closed \cite{wineland}. However, a much larger separation between the
ions or a shorter detection time are necessary for a loophole-free
test of a Bell inequality. For example, a realistic 50 $\mu$s
detection time will require 15 km separation between the ions
\cite{david2}. This may be difficult, since absorption of 369.5 nm
light in the optical fiber is relatively large ($\simeq 0.2$
dB/m). Therefore, frequency conversion \cite{tanzilli} or free space
light transmission may be an alternative solution. Entanglement over
larger distances could also be generated using the quantum repeater
protocol \cite{briegel}. Remote entanglement of atomic qubits is also
an important step towards the implementation of scalable quantum
information processing and scalable quantum communication networks.

This work is supported by the NSA and the IARPA under Army Research
Office contract, and the NSF Physics at the Information Frontier
Program.

\end{document}